 \documentclass[aps,floatfix,preprint,preprintnumbers,nofootinbib,superscriptaddress,natbib]{revtex4}

\usepackage{graphicx,float}% Include figure files
\usepackage{epsfig}		
\usepackage{dcolumn}% Align table columns on decimal point
\usepackage{bm}% bold math
\usepackage{subfigure}

\usepackage[all]{xy}
\usepackage{amsmath,upgreek}
\usepackage{amssymb}

\usepackage{pdfpages}
\usepackage{color}
\usepackage{graphicx,epstopdf}
\usepackage[colorlinks,hyperindex]{hyperref}

\def\0{\mbox{\tiny $0$}}
\def\1{\mbox{\tiny $1$}}
\def\2{\mbox{\tiny $2$}}
\def\3{\mbox{\tiny $3$}}
\def\4{\mbox{\tiny $4$}}
\def\5{\mbox{\tiny $5$}}
\def\6{\mbox{\tiny $6$}}
\def\7{\mbox{\tiny $7$}}
\def\8{\mbox{\tiny $8$}}
\def\9{\mbox{\tiny $9$}}

\def\f14{\mbox{\tiny $\frac{1}{4}$}}

\begin{document}

\title{CP symmetry and thermal effects on Dirac bi-spinor spin-parity local correlations}
\author{A. E. Bernardini}
\email{alexeb@ufscar.br}
\altaffiliation[On leave of absence from]{~Departamento de F\'{\i}sica, Universidade Federal de S\~ao Carlos, PO Box 676, 13565-905, S\~ao Carlos, SP, Brasil.}
\affiliation{Departamento de F\'isica e Astronomia, Faculdade de Ci\^{e}ncias da Universidade do Porto, Rua do Campo Alegre 687, 4169-007, Porto, Portugal.}
\author{V. A. S. V. Bittencourt}
\email{vbittencourt@df.ufscar.br}
\author{M. Blasone}
\email{blasone@sa.infn.it}
\affiliation{Dipartimento di Fisica, Universit\`a degli studi di Salerno, Via Giovanni Paolo II, 132 84084 Fisciano, Italy}
\altaffiliation{Also at: INFN Sezione di Napoli, Gruppo collegato di Salerno, Italy}

\date{\today}% It is always \today, today,
 % but any date may be explicitly specified
\renewcommand{\baselinestretch}{1.22}
\begin{abstract}
Intrinsic quantum correlations supported by the $SU(2)\otimes SU(2)$ structure of the Dirac equation used to describe particle/antiparticle states, optical ion traps and bilayer graphene are investigated and connected to the description of local properties of Dirac bi-spinors. For quantum states driven by Dirac-like Hamiltonians, quantum entanglement and geometric discord between spin and parity degrees of freedom - sometimes mapped into equivalent low energy internal degrees of freedom - are obtained. Such \textit{spin-parity} quantum correlations and the corresponding nonlocal intrinsic structures of bi-spinor fermionic states can be classified in order to relate quantum observables to the (non)local behavior of these correlations. It is shown that free particle mixed states do not violate the Clauser-Horne-Shymony-Holt inequality: the correlations in such mixed bi-spinors, although quantum, can be reproduced by a suitable local hidden variable model. Additionally, the effects due to a non-minimal coupling to a homogeneous magnetic field, and to the inclusion of thermal effects are evaluated, and quantum correlations of associated quantum mixtures and of the thermal states are all quantified.The above-mentioned correlation quantifiers are then used to measure the influence of CP transformations on \textit{spin-parity} quantum correlations, and our results show that quantum entanglement is invariant under CP transformations, although the geometric discord is highly sensitive to the CP symmetry.
\end{abstract}

%\pacs{03.65.Pm, 11.30.Er, 03.65.Ud}
\keywords{Dirac Equation - Bell Inequalities - Local Hidden Variables - Quantum Mechanics - Chirality - Parity}
\date{\today}
\maketitle

\section{Introduction}

Since the Einstein, Podolsky and Rosen \cite{EPR} controversial conclusion that quantum mechanics was not complete and that some underlying (hidden) variable would be necessary for a complete physical description of reality, mathematical and physical implications of a {\em (local) hidden variable theory} related to the phenomenon of quantum entanglement have been extensively investigated \cite{Nature01, Nature02, Rev01, Brunner, Rev03}.
The so-called hidden variable mechanism should support the interpretation of quantum mechanics as to account for the probabilistic features driven by some kind of inaccessible variable theory. A local hidden variable (LHV) theory adds the local realism as a requirement to rule out from the theory any kind of instantaneous non-causal measurement events.

According to the Bell's theorem \cite{Bell1}, some sets of LHV's cannot reproduce the quantum measurement correlations predicted by quantum mechanics.
The novel element introduced by Bell \cite{Bell1} was that quantum correlations are essentially nonlocal.
From the perspective of quantum entanglement \cite{Bengtsson,Peres,Simon,Adesso}, it states that separated particles can instantaneously share common properties and respond to quantum measurements as if they were a single particle.
This provides the setup for the analysis of the hidden variable scenario, leading for instance to Bell's inequalities \cite{Bell1,Banaszek,Brunner}. In such a context, the Bell inequality derived by Clauser, Horne, Shimony and Holt \cite{CHSH} -- the CHSH inequality -- has been used as a tool for testing locality in quantum systems. Through several setups, for example, in the characterization of superconducting qubits \cite{CHSH001}, in the architecture of trapped atoms \cite{CHSH002}, in the manipulation of quantum information protocols \cite{Brunner, Rev03}, or even in connection with noncomutative effects \cite{Bastos2016}, the CHSH inequality has been supported by measurements of observable quantities.

%Paragraph about relativistic quantum information - contextualization and general aspects
Bell's inequality has also been the ground for discussing quantum correlations in relativistic setups \cite{rel01, rel02, rel03, rel04, rel05, rel06}. Bell type correlations were firstly contextualized in a relativistic framework for discuss the concept of a relativistic center-of-mass \cite{rel01}, and since then the relation between Bell's inequality and transformation properties of quantum correlations was intensively studied \cite{rel02, rel03, rel04, rel05, rel05A, rel06}. For instance, when a pair of spins with different momenta is subjected to a Lorentz boost, the degree of violation of Bell's inequality is degraded \cite{rel01, rel02, rel03}, although for suitable conditions an anomalous behavior with local maximum of CHSH correlation, was observed \cite{rel05}. A complete covariant setup was described for states constructed as a pair of Dirac particles with a more complex behavior of CHSH spin-spin correlations under frame transformations \cite{rel04}. This relativistic effects on non-local spin correlations is due to the action of Lorentz boosts in spin states through a momentum dependent rotation, the Wigner rotation \cite{wigner}, and is the basis of many results in the fruitful field of relativistic quantum information \cite{revrel}. The behavior of spin-spin and spin-momentum entanglement under relativistic transformations has also been addressed \cite{rel06, rel07, rel08, rel09} and parallel to the discussion of transformation properties of quantum correlations, the very definition of relativistic spin operator has also been addressed \cite{rel05A, spin01, spin02, spin03, spin04, rel10}. In this fundamental context, a spin density matrix exhibiting covariant observable properties can be constructed with properly defined solutions of the Dirac equation \cite{rel10, spinorFW}.

The question to be posed in this manuscript is concerned with the quantification of the correspondence between (non)locality and quantum correlation aspects of the intrinsic structure of spin-$1/2$ particles described by relativistic bi-spinors, namely the (interacting) Dirac equation solutions.
Relativistic(-like) quantum systems has been preliminary considered under the perspective of quantum information theory \cite{Relinfo001, Relinfo002, Relinfo003} in a framework to test fundamental features of the relativistic quantum mechanics, in particular, those related to the computation of quantum correlations and their (covariant) transformation properties \cite{Relinfo002,Relinfo003}.
Supported by a $SU(2) \otimes SU(2)$ group structure \cite{SU202,MeuPRA}, Dirac bi-spinors exhibit a complete entanglement profile driven by two internal degrees of freedom -- the {\em spin} and the {\em intrinsic parity/chirality} \cite{Barrier,SU202} -- which provides an overall classification of the informational content of such Dirac-like structures \cite{SU202,MeuPRA}.

As a matter of fact, global potentials which eventually modify the free-particle Dirac dynamics also affect the above-mentioned intrinsic bi-spinor correlations \cite{SU202,MeuPRA} and bring up quantum correlation phenomenological aspects which can be identified not only in particle physics scenarios \cite{Scripta,Neutrinos01,Neutrinos02} but also in condensed matter \cite{MeuPRB} and quantum optical \cite{MeuPRA,JPB} relativistic-like mapped systems.
Given that all these systems are driven by Dirac-like equation solutions which -- through the $SU(2) \otimes SU(2)$ algebra -- intrinsically preserve the special relativity covariant structure, identifying and computing the quantifiers of quantum locality is by itself a relevant issue.

By following a global classification under Poincar\'{e} transformations, external field contributions to the Dirac dynamics are classified according to their (pseudo)scalar, (pseudo)vector and (pseudo)tensor characteristics \cite{Thaller}. A full Dirac Hamiltonian should read \cite{SU202, Thaller}
\begin{eqnarray}
\label{E04T}
\hat{H} &=& A^0(x)\,\hat{I}_4+ \hat{\beta}( m + \phi_S (x) ) + \hat{\bm{\alpha}} \cdot (\hat{\bm{p}} - \bm{A}(x)) + i \hat{\beta} \hat{\gamma}_5 \, \mu(x) - \hat{\gamma}_5 \, q(x) + \hat{\gamma}_5\, \hat{\bm{\alpha}}\cdot\bm{W}(x) \nonumber \\
&+& i \hat{\bm{\gamma}} \cdot [ \chi_a \bm{B}(x) + \kappa_a\, \bm{E}(x) \,] + \hat{\gamma}_5 \hat{\bm{\gamma}}\cdot[\kappa_a\, \bm{B}(x) - \chi_a \bm{E}(x) \,],
\end{eqnarray}
where bold variables denote vectors, with $a = \vert \bm{a} \vert = \sqrt{\bm{a} \cdot \bm{a}}$, hats ``$~\hat{}~$'' denote operators,
 $\hat{\beta}$ and $\hat{\bm{\alpha}} = \{\hat{\alpha}_x, \, \hat{\alpha}_y, \, \hat{\alpha}_z \}$ are the Dirac matrices that satisfies the anti-commuting relations
$\{\hat{\alpha}_i, \hat{\alpha}_j\} = 2 \, \delta_{ij} \, \hat{I}_4$, and
$\{\hat{\alpha}_i, \hat{\beta}\} =0$, with $i,j = x,y,z$, and
$\hat{\beta}^2 = \hat{I}_4$ (where $\hat{I}_N$ denotes the $N$-dim identity operator), all
with $\hat{\bm{\gamma}} = \hat{\beta} \hat{\bm{\alpha}}$, $\hat{\gamma}_5 = -i \, \hat{\alpha}_x \, \hat{\alpha}_y \, \hat{\alpha}_z$ (and, finally, with $\hbar$ and $c$ set equal to unity).
The bi-spinor quantum correlation structure driven by $\hat{H}$ is naturally obtained \cite{SU202,MeuPRA,Scripta} by identifying the $SU(2) \otimes SU(2)$ algebra operators with
\begin{equation}
\hat{\bm{\alpha}} = \hat{\sigma}_x \otimes \hat{\bm{\sigma}} \equiv \left[ \begin{array}{rr} 0 & \hat{\bm{\sigma}} \\ \hat{\bm{\sigma}} & 0 \end{array}\right],
\qquad \mbox{and} \qquad
\hat{\beta} = \hat{\sigma}_z \otimes \hat{I}_2 \equiv \left[ \begin{array}{rr} \hat{I}_2 & 0 \\ 0 & - \hat{I}_2 \end{array} \right],
\label{AAA}
\end{equation}
where $\hat{\bm{\sigma}}$ is the Pauli matrix vector representation that reflects the $\mbox{SU}(2)$ structure that leads to the interpretation of the Dirac quantum mechanics as an information theory for particles and fields. In this framework, the Dirac equation solutions are described by two-qubit states encoded by the Dirac continuous variable bi-spinor structure.
The simplest scenario corresponds to that one of Dirac bi-spinor eigenstates of the free particle Hamiltonian given in terms of two-qubit operators, ${H}_{D}={\hat{\sigma}}_{x}^{\left( 1\right) }\otimes \left(
{\bm p}\cdot {{\hat{\mbox{\boldmath$\sigma$}}}}^{\left( 2\right) }\right) +m \,
{\hat{\sigma} }_{z}^{\left( 1\right)}\otimes {I}^{(2)}_{2}$, written in terms of a sum of direct products describing {\em spin-parity} entangled states as
\begin{eqnarray}
\label{eqsB02}
\lefteqn{\left\vert \Psi ^{s}({\bm p},\,t)\right\rangle=
e^{i(-1)^{s}\,E_{p}\,t}\left\vert \psi ^{s}({\bm p})\right\rangle
}\\
&&= e^{i(-1)^{s}\,E_{p}\,t} N_{s}\left( p\right) \notag
 \left[ \left\vert
+\right\rangle _{1}\otimes \left\vert u({\bm p})\right\rangle _{2}+\left(
\frac{p}{E_{p}+(-1)^{s+1}m}\right) |-\rangle _{1}\,\otimes \left( {\bm p}
\cdot {\hat{\mbox{\boldmath$\sigma$} }}^{\left( 2\right) }\left\vert u(\bm{p}
)\right\rangle _{2}\right) \right],
\end{eqnarray}
where $s = 0,\, 1$ stands for particle/antiparticle associated frequencies, and the {\em spin} $1/2$ characteristic is obtained from $\left\vert u(\mathbf{p})\right\rangle_s$
\footnote{The state vector $\left\vert u(\mathbf{p})\right\rangle _{\2}$ is a bi-spinor that describes the dynamics of a fermion in momentum representation coupled to its {\em spin}: for the-qubit $1$, the {\em kets} $\left\vert + \right\rangle _{\1}$ and $ \vert - \rangle _{\1}$ are identified as the intrinsic parity eigenstates of the fermion and obey the orthonormalization relations $\left\langle \pm |\pm (\mp )\right\rangle _{\1}=1(0)$, as to give $\left\langle \psi ^{s}(\mathbf{p},\,t)|\psi ^{s}(\mathbf{p},\,t)\right\rangle =\,\,_{\2}\langle u(\mathbf{p})|u(\mathbf{p})\rangle _{\2}$, with
$$N_{s}(p)=\frac{1}{\sqrt{2}}\left( 1+(-1)^{s+1}\frac{m}{E_{p}}\right) ^{\1/\2}.
$$ }.
Dirac bi-spinors and {\em gamma matrices} represent the direct product between the internal degrees of freedom of spin and parity associated to a {\em spin} $1/2$ massive fermion.

The above structure, when extended to interacting systems as those ones described by \eqref{E04T}, supports the investigation of (non)local profiles of the quantum correlations -- as prescribed by the CHSH inequality -- for pure and mixed states of Dirac equation solutions describing fermionic particles.
Considering that the relevant {\em spin-parity} quantum correlation quantifiers are identified by the negativity (for entanglement, defined through Peres separability criterion) \cite{negativity01, negativity02}, and by the geometric discord \cite{geometricdiscordref} (for incremental correlations), the (non)local profile of quantum correlated states can be suitably computed from the inner parameters of the bi-spinor fermionic state \cite{bellquantifier}.
Our analysis shall be concerned with two relevant scenarios in the particle physics context: the free particle one, and the one for the neutral particle non-minimally coupled to a constant magnetic field. In particular, the latter case is also considered in a scenario which admits the inclusion of temperature effects on quantum correlations of the thermal state.

Besides the influence of thermal effects on the characterization of {\em spin-parity} quantum correlations, our analysis is motivated by the understanding of how the quantum locality can be affected by CP symmetry transformations. Given the relevance of the CP symmetry on symmetric fermion-antifermion processes in particle physics, any non-symmetric correspondence between CP properties and the quantum correlation quantifiers can be identified as a meter of CP violation.
In several scenarios of particle physics \cite{Hal84,Gri87,Giunti} which involve, for instance, neutrino and meson production and propagation, leptogenesis, and the overall pattern of weak interactions, the notwithstanding observation of CP violation -- explored by experiments in the lab -- provides a uniquely subtle link between the inner quantum space, which involve the {\em spin-parity} structure intrinsic to fermionic Dirac Hamiltonians, and the outer space, which is indeed phenomenologically connected to the CP violation observation.
Due to the relevance of CP symmetries in such scenarios of particle physics, its effects onto locality, entanglement and quantum correlation quantifiers are investigated herein.

The manuscript is thus structured as follows.
A brief review about the Dirac Hamiltonian system written as a composite quantum system is presented in section II. It includes the theoretical tools for obtaining the quantifiers for quantum entanglement, quantum correlations and CHSH inequalities, all specialized to the context of computing the nonlocal properties of Dirac bi-spinors.
In section III, the results for the negativity, the geometric discord and the Bell-CHSH function are obtained for some sets of mixed Dirac states.
The results include the maximal violation of the CHSH inequality as a probe for quantum nonlocality and how they are affected by the inclusion of thermal effects. The role of CP transformations onto the intrinsic quantum correlations, which shows how the local characteristics of Dirac bi-spinors are affected is discussed in section IV.
Our final conclusions are drawn in section V as to point to that, although minimal electromagnetic couplings, and non-minimal coupling to electric fields, do not produce quantum correlational CP violation effects, the non-minimal couplings with magnetic fields -- as those which typically modify the content of electroweak interactions -- indeed affect the quantum correlations driven by the geometric discord under CP transformations, in a kind of CP symmetry breaking.

\section{Nonlocal properties of Dirac bi-spinors}

Driven by the underlying $SU(2) \otimes SU(2)$ structure, the full Dirac Hamiltonian from (\ref{E04T}) can be rewritten as
\begin{eqnarray}
\label{fullhamiltonian}
\hat{H} &=& (\hat{\sigma}^{(1)}_z \otimes \hat{I}^{(2)})\, m + (\hat{\sigma}^{(1)}_x \otimes \hat{\bm{\sigma}}^{(2)}) \cdot \hat{\bm{p}}\nonumber \\
&&
\quad - (\hat{\sigma}_y ^{(1)} \otimes \hat{I}^{(2)})\mu(x) - (\hat{\sigma}_x^{(1)} \otimes \hat{I}^{(2)})\, q(x) + (\hat{I}^{(2)} \otimes \hat{\bm{\sigma}}^{(2)})\cdot\bm{W}(x)
\nonumber \\
&&
\quad\quad- (\hat{\sigma}_y^{(1)} \otimes \hat{\bm{\sigma}}^{(2)})\cdot [ \chi_a \bm{B}(x) + \kappa_a\, \bm{E}(x) \,] - (\hat{\sigma}_z^{(1)} \otimes \hat{\bm{\sigma}}^{(2)})\cdot[\kappa_a\, \bm{B}(x) - \chi_a \bm{E}(x) \,],\,
\end{eqnarray}
where the constant quadrivector components, $A_0$ and $\bm{A}$, have been dropped from the notation since they can be respectively absorbed by regular energy and momentum shifts.
The eigenstates of $\hat{H}$, with defined quantum numbers of parity and spin, describe a composite quantum system belonging to a composite Hilbert space, $\mathcal{H} = \mathcal{H}_P \otimes \mathcal{H}_S$, such that each subsystem is associated to a Hilbert space of dimension $2$.
The two corresponding subsystems in (\ref{fullhamiltonian}) are labeled with the superscripts $1$ and $2$ respectively associated to the degrees of freedom of intrinsic parity $P$ and the spin polarization $S$: the Dirac Hamiltonian is expressed as a two-qubit operator, and its eigenstates are two-qubit states.

As a consequence of the superposition principle for states describing composite quantum systems,
such two-qubit states can exhibit quantum entanglement with respective definition related to the concept of separability. It is better engendered in terms of an associated density matrix representation.
In fact, a generic two-qubit state can be written as
\begin{equation}
\label{decomposition}
\rho = \frac{1}{4} \left[\hat{I} + (\hat{\bm{\sigma}}^{(1)} \otimes \hat{I}_2^{(2)}) \cdot \bm{a}_1 + (\hat{I}_2^{(1)} \otimes \hat{\bm{\sigma}}^{(2)})\cdot \bm{a}_2 + \displaystyle \sum_{i,j = \{x,y,z\}} (\hat{\sigma}_i \otimes \hat{\sigma}_j) \, t_{ij} \right],
\end{equation}
where $\bm{a}_{1 (2)}$ are the so-called Bloch vectors and the elements $t_{ij}$ are identified as the components of a $3 \times 3$ correlation matrix, $T$.
A bipartite state $\rho \in \mathcal{H}_1 \otimes \mathcal{H}_2$ is separable if it can be written as
\begin{equation}
\rho = \displaystyle \sum_{i} c_i \, \rho_i^{(1)} \otimes \rho_i^{(2)},
\end{equation}
with $c_i \ge 0$, $\sum_i c_i = 1$, $\rho_i^{(1)} \in \mathcal{H}_1$ and $\rho_i^{(2)} \in \mathcal{H}_2$. If $\rho$ is not a separable state, then it is entangled. Entangled and separable states can be distinguished by the Peres criterion, which asserts that if a state $\rho$ is separable then its partial transposition with respect to any of the subsystem $\rho^{T_i}$ is also a valid density matrix.
From this criterion it is possible to demonstrate that $\rho$ is separable if $\rho^{T_i}$ has only positive eigenvalues. With respect to some fixed basis of the composite Hilbert space $\mathcal{H}$, $\{\vert \mu_i \rangle \otimes \vert \nu_j \rangle \}$ (with $\vert \mu_i \rangle \in \mathcal{H}_1$ and $\vert \nu_i \rangle \in \mathcal{H}_2$), the matrix elements of the partial transpose with respect to the first subsystem $\rho^{T_1}$ are given by
\begin{equation}
\langle \mu_i \vert \otimes \langle \nu_j \vert \rho^{T_1} \vert \mu_k \rangle \otimes \vert \nu_l \rangle = \langle \mu_k \vert \otimes \langle \nu_j \vert \, \rho \, \vert \mu_i \rangle \otimes \vert \nu_l \rangle.
\end{equation}
In terms of the Fano decomposition (\ref{decomposition}), the partial transposition with respect to the first qubit reads
\begin{equation}
\rho^{T_1} = \frac{1}{4} \left[ (\hat{\bm{\sigma}}^{(1)} \otimes \hat{I}_2^{(2)}) \cdot \bm{b}_1 + (\hat{I}_2^{(1)} \otimes \hat{\bm{\sigma}}^{(2)})\cdot \bm{b}_2 + \displaystyle \sum_{i,j = \{x,y,z\}} (\hat{\sigma}_i \otimes \hat{\sigma}_j) \, q_{ij} \right],
\end{equation}
where $\bm{b}_1 = (a_{1x}, - a_{1y}, a_{1z})$, $\bm{b}_2 = \bm{a}_2$, and
\begin{eqnarray}
Q \equiv \{q_{ij}\}&=&\left[ \begin{array}{rrr} t_{xx} & t_{xy} & t_{xz} \\ - t_{yx} & - t_{yy} & - t_{yz} \\ t_{zx} & t_{zy} & t_{zz} \end{array} \right],\nonumber
\end{eqnarray}
written in terms of the original Bloch vector and correlation matrix components, $a_{1j}$ and $t_{ij}$, with $i,j=x,\,y,\,z$.

According to the Peres separability criterion, the entanglement quantifier of a quantum state $\rho$ is identified by the negativity, $\mathcal{N}$, given by
\begin{equation}
\label{negativity}
\mathcal{N}[\rho] = \displaystyle{\sum}_{i} \vert \mu_i \vert - 1,
\end{equation}
where $\mu_i$ are the eigenvalues of the matrix $\rho^{T_i}$. If the state is entangled, at least one of the $\mu_i$ elements will be negative, and thus $\mathcal{N}[\rho] > 0$. On the other hand, for separable states all eigenvalues $\mu_i$ will be positive and thus $\mathcal{N}[\rho] = 0$. Moreover, one has $\mathcal{N}[\rho] = 1$ for the maximally entangled states.

Otherwise, mixed states can display quantum correlations even when they are not entangled \cite{quantumcorre}, but a complete characterization of them is still considered a partially open problem. Quantum discord, for instance, was the first measure proposed to quantify this kind of non-classical correlations \cite{quantumdiscord}.
It has been applied, for example, in the study of phase transitions \cite{Discord01}, as well as in connection with quantum information theory as a resource for quantum computation \cite{Discord02}. Although quantum discord encodes all non-classical correlations for a given state, its computation involves an optimization process over a set of all possible projectors into one of the subsystems.
From a computational point of view, it corresponds to a NP-hard problem which may not have an analytical form, even for the simplest two-qubit cases \cite{discordNPhard}. This issue can be circumvented by adopting a geometric measurement called geometric discord, $\mathcal{D}$, defined as the minimal Hilbert-Schmidt distance between a given state and the set of zero quantum discord states \cite{geometricdiscordref}.
The geometric discord contains the same information about the quantum correlation encoded by the quantum discord, with the advantage of having an analytical form for two-qubit states. It reads
\begin{equation}
\label{geometricdiscord}
\mathcal{D}_{1(2)}[\rho] = \frac{1}{4}(a_{1(2)}^2 + \vert \vert T \vert \vert ^2 - k_{\mbox{\tiny max}}),
\end{equation}
where $a_{1(2)}^2 = \bm{a}_{1 (2)}\cdot \bm{a}_{1(2)}$, $\vert \vert T \vert \vert ^2 = \mbox{Tr}[T \, T^T]$, and $k_{\mbox{\tiny max}}$ is the largest eigenvalue of the matrix $\bm{a}_{1 (2)} \bm{a}^T_{1(2)} + T T^T$. In particular, for any two-qubit state one has $\mathcal{D}[\rho] \ge (\mathcal{N}[\rho])^2$.

Besides the geometric discord, for mixed states, additional quantum correlations associated to the concept of nonlocality are still not completely well-established.
The engendering of nonlocality quantifiers was initially proposed for investigating the physical consequences of entanglement in setups involving the measurement of outcome probabilities in quantum measurement experiments.
From a theoretical point of view, it has been expressed in terms of Bell inequalities, which are indeed a family of inequalities satisfying very suitable requirements.

One of the most important Bell inequalities is the CHSH inequality, which provides a way to test the LHV model associated to a supposed correlation.
To clear up the meaning of the CHSH inequality, one can consider a setup where two dichotomic variables only admit the discrete values $+1$ and $-1$. The dichotomic variables $A_1$ and $A_2$ are measured in the subsystem $1$ and those $B_1$ and $B_2$ are measured in the subsystem $2$. The LHV theory imposes that the joint expectation values $E(A_i, B_j)$ must satisfy the inequality
\begin{equation}
\vert E(\hat{A}_1, \hat{B}_1) + E(\hat{A}_1, \hat{B}_2) + E(\hat{A}_2, \hat{B}_1) - E(\hat{A}_2, \hat{B}_2) \vert \le 2.
\end{equation}
For states that do not admit a LHV model, the inequality is violated. In the case of a two-qubit system, the CHSH inequality is evaluated through the mean value of the operator
\begin{equation}
\label{bell01}
\hat{B} = \hat{\bm{A}}_1 \otimes \hat{\bm{B_1}} + \hat{\bm{A}}_1 \otimes \hat{\bm{B_2}} + \hat{\bm{A}}_2 \otimes \hat{\bm{B_1}} - \hat{\bm{A}}_2 \otimes \hat{\bm{B_2}},
\end{equation}
where $\hat{\bm{A}}_i = \bm{A}_i \cdot \hat{\bm{\sigma}}^{(1)}$ and $\hat{\bm{B}}_i = \bm{B}_i \cdot \hat{\bm{\sigma}}^{(2)}$, with $i = 1,\,2$. A state $\rho$ admits a LHV model if, for at least one choice of the vectors $\bm{A}_i$ and $\bm{B}_i$, one has $\vert \, \mbox{Tr}[\hat{B} \rho] \, \vert \le 2$. Therefore, to measure the locality property of the correlations between two subsystems in a two-qubit state, one needs to find the vectors $\bm{A}_i$ and $\bm{B}_i$ which maximize the mean value of $\mbox{Tr}[\hat{B} \rho]$. If the mean value is larger than $2$, the state is said to be nonlocal, and the correlations in such states can not be described by a LHV model.

For an arbitrary two-qubit state $\rho$ as given by (\ref{decomposition}), the maximum possible mean value of (\ref{bell01}) can be computed in terms of the correlation matrix of the state through the formula \cite{bellquantifier}
\begin{equation}
\mbox{max}_{\bm{A}_i, \bm{B}_i} \vert \mbox{Tr}[\hat{B} \rho] \vert = 2 \sqrt{t_{1} + t_2},
\end{equation}
where $t_1$ and $t_2$ are the largest eigenvalues of the matrix $T^T T$.
By identifying a locality quantifier with $M[\rho] = t_{1} + t_2$, a sufficient criterion for the violation of the CHSH inequality reads
\begin{equation}
M[\rho] > 1,
\end{equation}
which can be rewritten in terms of the re-defined Bell function,
\begin{equation}
\label{bellfunction}
\mathcal{B}[\rho] = M[\rho] - 1,
\end{equation}
as $\mathcal{B}[\rho] > 0$.
It can also be demonstrated that its maximum possible value, $\mbox{max}_{\bm{A}_i, \bm{B}_i} \vert \mbox{Tr}[\hat{B} \rho] \vert$ is equal to $2 \sqrt{2}$, which is only attainable by pure maximally entangled states. Therefore, $\mathcal{B}[\rho]$ quantifies the degree at which the CHSH inequality is violated.

Once the above quantifiers have been established, one can turn attentions to the analysis of bi-spinor states with the two degrees of freedom intrinsic parity and spin. The (total intrinsic plus kinematic) parity operator $\hat{\mathcal{P}}$ acts on the direct product of free particle states (\ref{eqsB02}) $\vert \pm \rangle _{\1}\otimes \vert u(\mathbf{p})\rangle_{\2}$ in the form of $$\hat{\mathcal{P}}\left( \vert \pm \rangle _{\1}\otimes \vert u(\mathbf{p})\rangle _{\2}\right) =\pm \left( \vert \pm \rangle_{\1}\otimes \vert u(-\mathbf{p})\rangle_{\2}\right),$$ from where one identifies the intrinsic parity operator $\hat{P}$ as 
\begin{equation}
\hat{P} = \hat{\beta} = \hat{\sigma}_z \otimes \hat{I},
\end{equation}
acting only on the first qubit of the spin-parity decomposition of a given bi-spinor. On the other hand, due to its intrinsic relation with group-theory concepts and its importance for implementations of information protocols, several propositions of spin operators for relativistic particles were proposed in the literature \cite{spin01, spin02, spin03, spin03A}. The choice of spin operator used to compute Bell's inequality in two particle setup influence the degree of non-locality measured \cite{rel01, rel05A, rel06}, affecting also possible definitions of position operator in the relativistic quantum mechanics framework \cite{rel06} and changing predictions of observables in the presence of electromagnetic fields \cite{spin04}. The simpler spin operator in the context of Dirac equation solutions is the so called Pauli spin operator
\begin{equation}
\label{spinop}
\hat{\bm{S}}_{Pauli} = \frac{1}{2} \hat{\bm{\Sigma}} = \frac{1}{2} \hat{I} \otimes \hat{\bm{\sigma}},
\end{equation}
which acts on the basis vector used to write the two qubit form of the Hamiltonian (\ref{decomposition}) as $$2 \, \hat{\bm{S}}_{Pauli} \, \vert \pm \rangle_1 \otimes \vert \pm \rangle_2 = \vert \pm \rangle_1 \otimes( \,\pm \vert \pm \rangle_2 + (1 \pm i)\vert \mp \rangle_2 \,),$$
and the maximal degree of violation of Bell's inequality given by (\ref{bellfunction}) is obtained through a maximization process over all possible spin polarizations. It is worth to mention that other possible spin operator is the Fouldy-Wouthuysen (FW) spin operator \cite{FW}, which additionally to (\ref{spinop}) has a term dependent on the momentum of the particle. The mean value of spin as measured by FW spin operator is a constant of the free Dirac dynamics and the eigenstates of FW operator were used to discuss transformation properties of spin-spin entanglement \cite{spinorFW} as well as in connection with the definition of position operator in relativistic quantum mechanics \cite{rel10}.

In the present single particle spin-parity setup, for pure states driven by the Hamiltonian which only includes tensor and pseudo-tensor interactions, the results for nonlocality simply follow the results for entanglement, i.e. any entangled pure state violates the CHSH inequality and, therefore, its correlation cannot be described through a LHV.
The mixed state scenario is more involved: the relation between entanglement and nonlocality is not fully understood even for the two-qubit states. Even for free particle Dirac Hamiltonian structures, mixed states exhibit an intrinsic {\em spin-parity} correlation profile where the correspondence between entanglement and nonlocality changes according to the kinematic regime.
The complexity is enhanced when one includes tensor and pseudotensor external potentials into the Dirac Hamiltonian.
%Notwithstanding its connection with weak interactions in particle physics, such a (pseudo)tensor interacting framework is still more relevant in mapping suitable Jaynes-Cummings interactions that describe four level trapped ion systems, providing a bridge to the study intrinsic correlation properties in the trapped ion optical setups.

\subsection{Free particle mixed helicity states}

From the free particle solutions described by \eqref{eqsB02}, helicity eigenstates, $\vert h_{\pm} \rangle$, can be identified by the action of the normalized projection operator $\hat{h} = \bm{\hat{\Sigma}}\cdot \bm{p}/p$ as $$\vert h_{\pm} \rangle = \frac{1 \pm \bm{\hat{\sigma}} \cdot(\bm{p}/p)}{\sqrt{2}} \vert \pm \rangle,$$ with eigenvalues $\pm 1$.
The corresponding positive frequency solutions written in terms of helicity eigenstates is thus given by
\begin{eqnarray}
\label{helictystate}
\vert \psi_{\pm} (\bm{x},t) \rangle = \sqrt{\frac{E_p - m}{2 E_p}} \left[\vert 1 \rangle \otimes \vert h_{\pm} \rangle \pm \sqrt{\frac{E_p + m}{E_p - m}}\vert 0 \rangle \otimes \vert h_{\pm} \rangle \right] \exp{[- i (E_p t - \bm{p} \cdot \bm{x})]},
\end{eqnarray}
which, by the factorization of $\vert h_{\pm} \rangle $, are evidently separable {\em spin-parity} states.
Quantum superpositions of $\vert \psi_+ \rangle$ and $\vert \psi_- \rangle$ are unavoidably entangled states. Such an entanglement content has been completely quantified in the context of planar (2D) diffusion problems \cite{Barrier} and it is potentially relevant for 2D condensed matter systems like graphene \cite{MeuPRB}.
For the planar diffusion of spin-polarized electrons, the potential step\footnote{With the step energy, $A^0$, commonly absorbed by the energy eigenvalues.} can generate entanglement for an initially incident separable state, and also destroy its quantum correlation even for a maximally entangled initial states in a kind of Brewster angle dependent entanglement production \cite{Barrier}.

\begin{figure}[H]
\centering
\includegraphics[width= 8 cm]{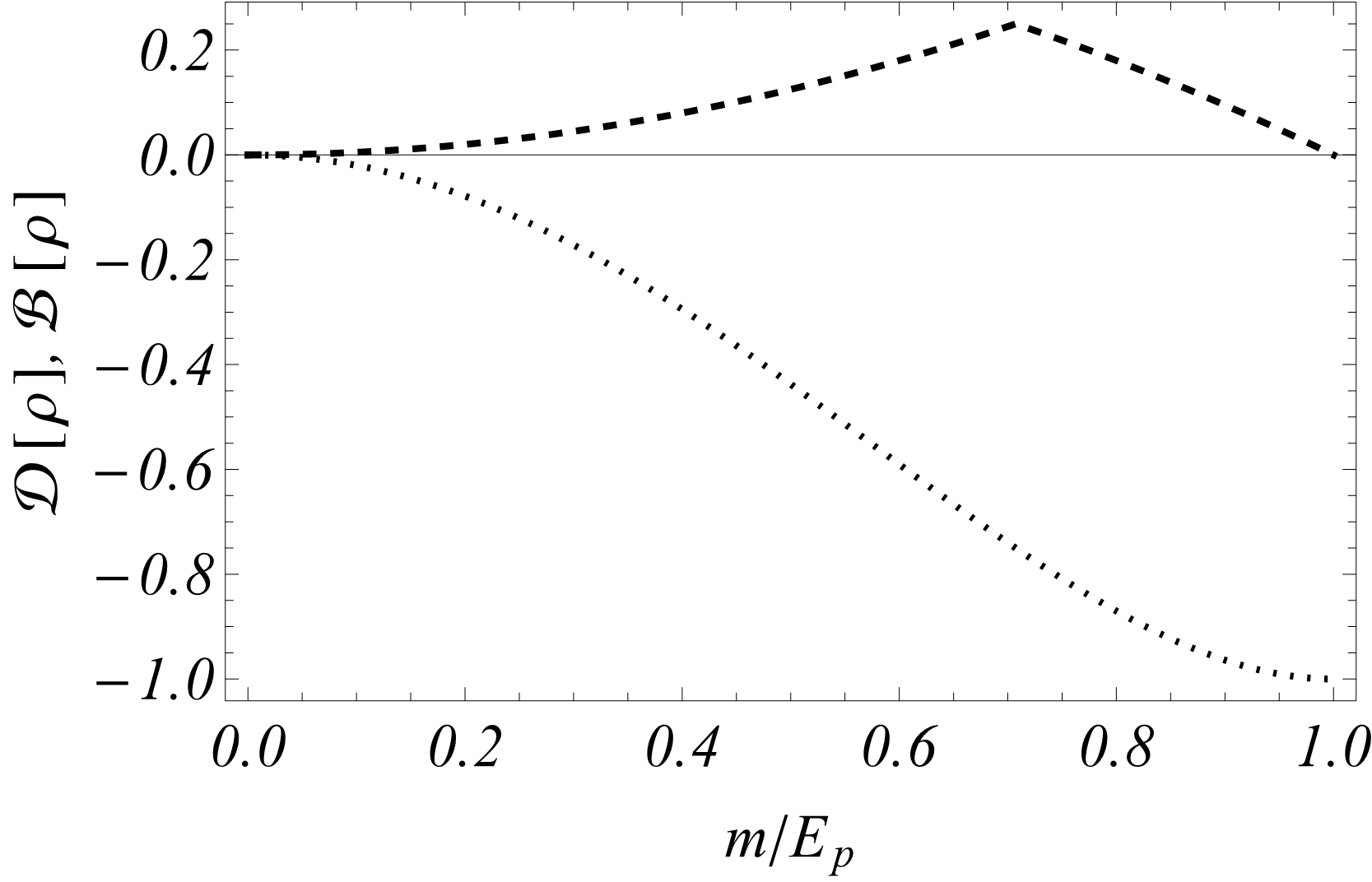}
\renewcommand{\baselinestretch}{1.0}
\caption{Geometric discord (dashed line) and the Bell function (dotted line) for the free particle maximal mixture of helicity (separable) states (\ref{freeparticle}) as function of $m/E_p$.
Although the geometric discord is not null (dashed lines), the {\em spin-parity} quantum correlational content of $\rho_{\mbox{\tiny free}}$ is always of local nature, as the Bell function (dotted lines) is always negative. The quantum correlations quantified by the geometric discord, in this case, can be reproduced by a LHV theory.}
\label{FigFreeParticle}
\end{figure}

Without loss of generality, the electron wave function propagation can be discussed in a frame with a unidimensional wave vector, $\bm{p} = p \hat{z}$. The positive energy mixed state $\rho_{\mbox{\tiny free}} = A \, \vert \psi_+ (z,t) \rangle \langle \psi_+ (z,t) \vert + (1 - A) \vert \psi_- (z,t) \rangle \langle \psi_- (z,t) \vert$ is then explicitly given by
\begin{eqnarray}
\label{freeparticle}
\rho_{\mbox{\tiny free}} &=& \left[ \begin{array}{cccc} (1-A) \frac{E^2_p - m^2}{2 E_p} & 0 & -(1 - A)\frac{\sqrt{E^2_p - m^2}}{2 E_p } & 0 \\ 0 & A \frac{E_p + m}{2 E_p} & 0 & A \frac{\sqrt{E^2_p - m^2}}{2 E_p} \\ -(1 - A)\frac{\sqrt{E^2_p - m^2}}{2 E_p} & 0 & (1-A) \frac{E_p - m}{2 E_p} & 0 \\ 0 & A \frac{\sqrt{E^2_p - m^2}}{2 E+p} & 0 & A \frac{E_p - m}{2 E_p}\end{array} \right],
\end{eqnarray}
from which one has $\rho_{\mbox{\tiny free}}^{T_1} = \rho_{\mbox{\tiny free}}$, such that $\rho_{\mbox{\tiny free}}$ is a separable state (c.f. the Peres criterion).
However, it exhibits quantum correlations of other nature, quantified by the geometric discord (\ref{geometricdiscord}), evaluated as
\begin{equation}
\label{discordfreeparticle}
\mathcal{D}[\rho_{\mbox{\tiny free}}] = \frac{1}{4} \left[\, \left(1 + \, (1- 2 A)^2 \frac{m^2}{E_p^2}\,\right) - \sqrt{\left(1 + \, (1- 2 A)^2 \frac{m^2}{E_p^2}\,\right)^2 - 4 \left(1 - \frac{m^2}{E_p^2}\right)\frac{m^2}{E_p^2}} \, \right],
\end{equation}
which vanishes only for $A = 0$ and for $m/E_p = 0$ or $1$ (i.e. for ultrarelativistic and nonrelativistic limits). In spite of their quantum nature, these correlations are local. In fact, for (\ref{freeparticle}) the quantifier (\ref{bellfunction}) is evaluated as
\begin{equation}
\mathcal{B}[\rho_{\mbox{\tiny free}}] = \left(1 - 4 A (1 - A) \frac{m^2}{E_p^2}\right)^2 - 1,
\end{equation}
and therefore, if $\mathcal{D}[\rho_{\mbox{\tiny free}}] \neq 0$, then $\mathcal{B}[\rho_{\mbox{\tiny free}}] < 0$ and (\ref{freeparticle}) does not violate the CHSH inequality for any set of parameters. The quantum correlations quantified by a non-vanishing $\mathcal{D}$ values in (\ref{discordfreeparticle}) are therefore local. Moreover, the geometric discord (\ref{discordfreeparticle}) is maximized for states with mass $m_{\mbox{\tiny max}}$ given by $$m_{\mbox{\tiny max}} = \sqrt{\frac{2 (1-A)}{5 -8 A + 4 A^2 }}.$$ For a maximal mixture between positive and negative helicity states, i.e. for $A = 0.5$, one has a cuspid point for $\mathcal{D}$ (the first derivative of is discontinuous) as depicted in Fig.~\ref{FigFreeParticle}.

\subsection{Inclusion of tensor and pseudotensor couplings}

The Dirac Hamiltonian with tensor and pseudotensor potentials
includes couplings with constant and homogeneous electric and magnetic fields, respectively, which naturally induces the intrinsic {\em spin-parity} entanglement.
The Hamiltonian for a neutral particle non-minamilly coupled with an external magnetic field $\bm{B}$ through an anomalous magnetic moment $\kappa_a$, and an exotic axial-vector interaction driven $\chi_a$ (included by completeness) is given by
\begin{equation}
\label{hamiltonian}
\hat{H} = \hat{\beta} \, m + \hat{\bm{\alpha}} \cdot \bm{p} + i \chi_a \hat{\bm{\gamma}} \cdot \bm{B} + \kappa_a \,\hat{\gamma}_5 \, \hat{\bm{\gamma}}\cdot \bm{B},
\end{equation}
with the underlying $SU(2) \otimes SU(2)$ structure expressed by
\begin{equation}\hat{H} = (\sigma_z^{(1)} \otimes \hat{I}^{(2)}) \, m + (\sigma_x^{(1)}\otimes \hat{\bm{\sigma}}^{(2)})\cdot \bm{p} + \kappa_a \, (\sigma_z^{(1)}\otimes \hat{\bm{\sigma}}^{(2)})\cdot \bm{B} - \chi_a (\sigma_y^{(1)}\otimes \hat{\bm{\sigma}}^{(2)}) \cdot \bm{B}.
\end{equation}

Much more relevant than a possible appeal to some neutrino/neutron non-minimal coupling interaction driven by $\hat{H}$ \cite{Neutrinos02}, the above related Hamiltonian eigenstates can also be simulated by a four-level trapped ion setup (used, for instance, to simulate the \textit{zitterbewegung} effect) through which the {\em spin-parity} entanglement of the corresponding Dirac equation solutions can be mapped into the entanglement between two ionic variables: one related to the total angular momentum and another one related to its projection onto the direction of a magnetic field used to lift the ionic degeneracy \cite{MeuPRA}.

By observing some algebraic properties of $\hat{H}$, from Eq.~\eqref{hamiltonian} one has
\begin{eqnarray}
\hat{H}^2 &=& c_1 \hat{I} + 2 \hat{\mathcal{O}}, \nonumber \\
\hat{\mathcal{O}}^2 &=& \frac{1}{4}\left( \hat{H}^2 - c_1 \hat{I}\right)^2 = c_2 \hat{I},
\end{eqnarray}
with
\begin{eqnarray}
c_1 &=& \frac{1}{4}\mbox{Tr}[\, \hat{H}\,] = p^2 + m^2 + B^2 (\kappa_a^2 + \chi_a^2)\nonumber \\
c_2 &=& \frac{1}{16} \mbox{Tr}[\,(\mbox{Tr}[\hat{H}^2] - \frac{1}{4}\mbox{Tr}[\hat{H}^2])] = m^2 \kappa_a^2 \, B^2 + (\kappa_a^2 + \chi_a^2)(\bm{p} \times \bm{B})^2,
\end{eqnarray}
and, of course,
\begin{eqnarray}
\hat{\mathcal{O}} &=& \frac{1}{2}\left(\hat{H}^2 - c_1 \hat{I}_4\right) = m \, \kappa_a \, \hat{\bm{\Sigma}}\cdot \bm{B} + \chi_a \hat{\beta} \, \hat{\bm{\Sigma}}\cdot (\bm{p} \times \bm{B}) - i \kappa_a \hat{\beta} \hat{\bm{\alpha}} \cdot (\bm{p} \times \bm{B}).
\end{eqnarray}

It provides the eigenvalues of \eqref{hamiltonian} as given by $\lambda_{n,s} = (-1)^n \sqrt{c_1 + 2 (-1)^s \sqrt{c_2}}$, which describe a non-degenerate energy spectrum (as $c_2 \neq 0$).
In this case, the density matrices associated to the eigenstates $\rho_{n,s}$ must be a third degree polynomial of (\ref{hamiltonian}) which, according to the algebraic strategy introduced in Ref.~\cite{SU202}, can be written as
\begin{equation}
\label{anomalouseigenstate}
\rho_{n,s} = \left[\hat{I}_4 + \frac{(-1)^n}{\vert \, \lambda_{n,s} \, \vert} \hat{H}\right] \left[ \hat{I}_4 + \frac{(-1)^s}{\vert \, \sqrt{c_2} \, \vert} \hat{\mathcal{O}} \right].
\end{equation}

For a generic mixed state given in terms of the eigenstates of (\ref{hamiltonian}),
\begin{equation}
\label{anomalousstate}
\rho =\displaystyle \sum_{n,s} A_{n,s} \, \rho_{n,s},
\end{equation}
with $\sum_{n,s} A_{n,s} = 1$ and $A_{n,s} >0$, the Bloch vectors and the elements of the correlation matrix are given by
\begin{eqnarray}
\label{parameters}
\bm{a}_1 &=& (- g_3 \, m \kappa_a \, \bm{p} \cdot \bm{B}, \, g_3 \, m \chi_a \kappa_a \, B^2, \, g_1 \, m + g_3\, m \kappa_a^2 \, B^2), \nonumber \\
\bm{a}_2 &=& g_3 \chi_a m \, \bm{\omega} - g_2 \, m \kappa_a \bm{B}, \nonumber \\
t_{x i} &=& g_3 (\chi_a^2 + \kappa_a^2) (\bm{B} \times \bm{\omega})_i, \nonumber \\
t_{y i} &=& g_3 \, \chi_a (\bm{p} \times \bm{\omega})_i - g_2 \, \kappa_a \omega_i - g_1 \chi_a B_i, \nonumber \\
t_{z i} &=& g_3 \, \kappa_a ((\bm{p} \times \bm{\omega})_i - m^2 B_i) + g_2 \, \chi_a \omega_i - g_1 \, \kappa_a B_i,
\end{eqnarray}
where $\bm{\omega} = \bm{p} \times \bm{B}$ and the coefficients $g_{1,2,3}$ are given by
\begin{eqnarray}
g_1 = \displaystyle \sum_{n,s}\frac{A_{n,s}}{\lambda_{n,s}}; \qquad g_2 = \frac{1}{\sqrt{c_2}} \displaystyle \sum_{n,s}(-1)^s A_{n,s}; \qquad g_3 = \frac{1}{\sqrt{c_2}}\displaystyle \sum_{n,s}\frac{(-1)^s A_{n,s}}{\lambda_{n,s}}.
\end{eqnarray}

By noticing the complexity of the above dependence on the manipulable variables, from now on, without loss of generality, one may have adopted $\bm{p} = p \, \bm{i}$ and $\bm{B} = B \cos{\theta} \, \bm{i} + B \sin{\theta} \, \bm{j}$. Since the values of $\theta$ does not interpretatively affect the results, it shall be used $\theta = \pi/4$.

\begin{figure}[H]
\includegraphics[width= 16.5 cm]{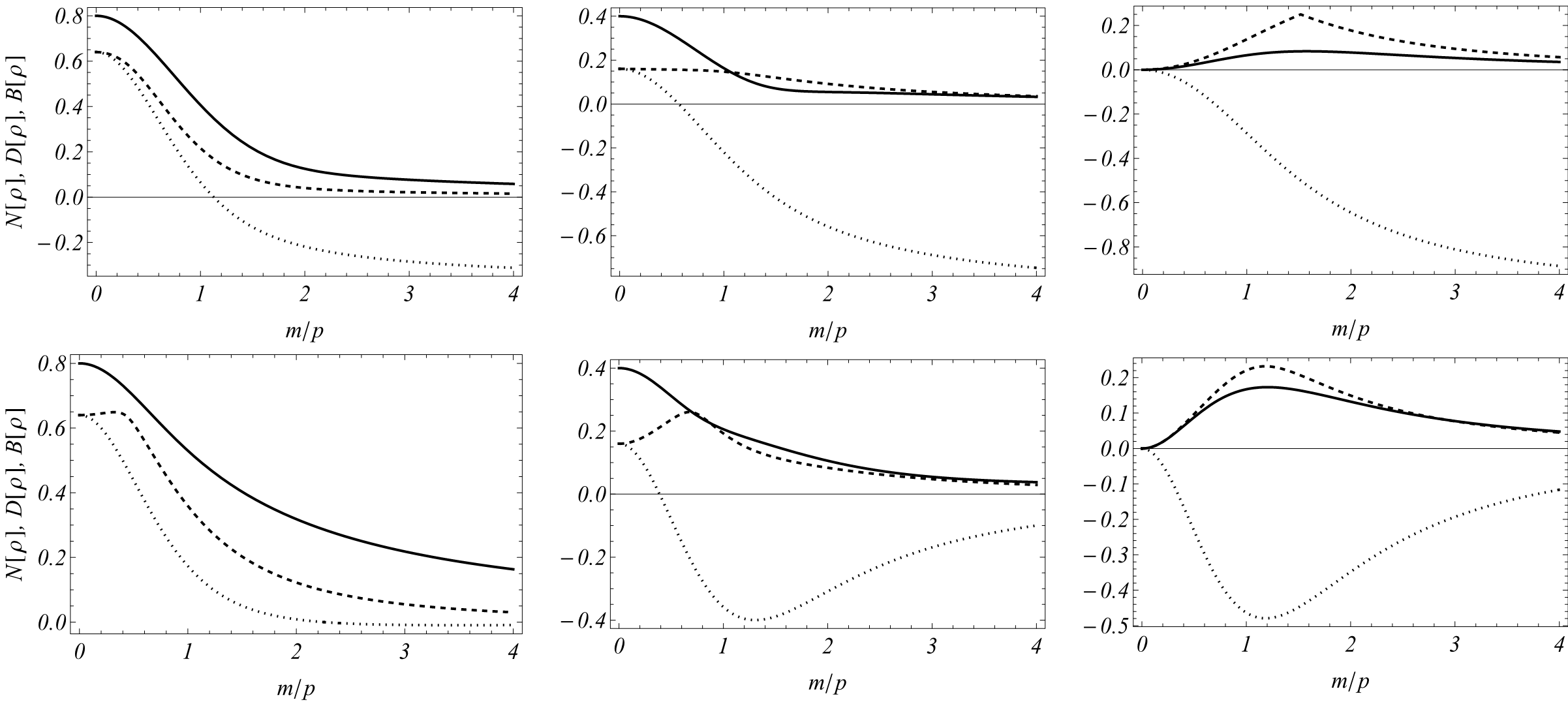}
\renewcommand{\baselinestretch}{1.0}
\caption{Negativity (solid line), geometric discord (dashed line) and Bell function (dotted line) as function of $m/p$ for mixtures (c.f. Eq.~(\ref{anomalousstate})) given by $A \rho_{00} + (1 - A) \rho_{01}$ (first row) and $A \rho_{00} + (1 - A) \rho_{11}$ (second row), for $A = 0.1$ (first column), $0.3$ (second column) and $0.5$ (third column). Plots are for $B/p =\chi_a = \kappa_a = 1$ and $\theta = \pi/4$. In the nonrelativistic limit ($m \gg p$) the state is local and separable. For $m \sim p$ the state is entangled but can be local or nonlocal depending on the value of $A$. For the maximal mixture, both states are local for any value of $m/p$. In this case, they are entangled and display quantum correlations concentrated around a maximum value. Superpositions involving positive and negative energy eigenstates increase the nonlocality with respect to mixed state of positively (or negatively) defined energies.}
\label{EigenstatesTensor}
\end{figure}

Fig.~ \ref{EigenstatesTensor} depicts the negativity $\mathcal{N} [\rho]$ (solid lines), the geometric discord $\mathcal{D}[\rho]$ (dashed lines) and the Bell-CHSH quantifier $\mathcal{B}[\rho]$ (dotted lines) for mixtures between positive energy states resumed by $A \rho_{00} + (1 - A) \rho_{01}$ (first row) and between positive and negative energy states resumed by $A \rho_{00} + (1 - A) \rho_{11}$ (second row) as function of $m/p$.
 For non-relativistic states, i.e. for $m \gg p$, the mixture is separable, local and does not display any type of quantum correlations. Otherwise, when the momentum and the mass are of the same magnitude, the state can be either nonlocal and entangled or local and entangled. As in the free particle case, for a maximal mixture, the state is local but displays quantum correlations concentrated around a maximal value, for which $\mathcal{D}[\rho]$ again presents a discontinuous first derivative (for the positive energy mixture). Mixed states engendered by positive and negative energy quantum superpositions exhibit an increasing nonlocal pattern, given that for such states $\mathcal{B}$ negatively increases. In particular, for $A = 0.1$, the mixture always violates the CHSH inequality.

\subsection{Inclusion of thermal effects}

In the present framework it is possible to study how the temperature affects the correlations.
In a simplified scenario where the system -- initially described as an arbitrary state -- interacts with a heat bath at the inverse temperature $\beta$, in the absence of time-dependent external fields it can be proved that the thermal state (or the Gibbs state) $$\rho_{\mbox{{\tiny Gibbs}}} = {e^{- \beta \, \hat{H}}}/{\mbox{Tr}[e^{- \beta \, \hat{H}}]},$$ is a stationary solution of the master equation describing the evolution of the system under the action of such an environment. Given that the quantum dynamical semigroup describing this open system evolution has the ergodic property, any initial state will evolve to the thermal state under the action of a heat bath \cite{breuer}.

\begin{figure}[H]
\includegraphics[width= 16.5 cm]{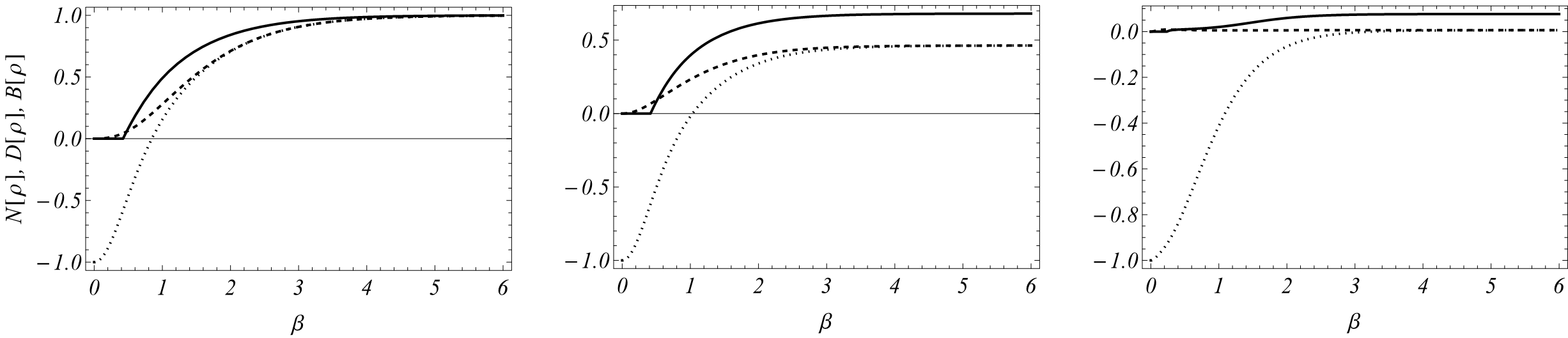}
\renewcommand{\baselinestretch}{1.0}
\caption{Negativity (solid line), geometric discord (dashed line) and Bell function (dotted line) as function of $\beta = 1/T$ for the thermal state, for $m/p = 0$ (first column), $1$ (second column) and $10$ (third column). All other parameters are in correspondence with Fig.~\ref{EigenstatesTensor}. For high temperatures, the state tends to the maximal mixture which does not exhibit any quantum correlation. As the temperature decreases, quantum correlations gradually increase. In particular, quantum entanglement suddenly appears at some precision specific correspondence between $T$ and $m/p$. For low temperatures, the quantum entanglement, quantum correlations and the locality quantifiers tend to a constant value (with a common value between Bell functions and geometric discord). In the limit of zero mass and low temperatures, the mixture tends to a maximally entangled state.}
\label{thermal01}
\end{figure}

The thermal state of (\ref{hamiltonian}) corresponds to a generic thermalized state given by
\begin{equation}
\label{thermal}
\rho_{\mbox{{\tiny Thermal}}} = \displaystyle
\left( \displaystyle \sum_{\{k,l\}} e^{-\beta \lambda_{k,l}}\right)^{-1} \sum_{\{n,s\}} {e^{- \beta \lambda_{n,s}}} \rho_{n,s},
\end{equation}
which corresponds to a generic mixture which has correlation properties depending on the temperature. In the high temperature limit ($\beta \rightarrow 0$), the thermal state tends to the maximally mixed state $\hat{I}/4$, and in the low temperature limit ($\beta \gg \lambda_{n,s}$) it tends to the lowest energy eigenstate.
For the thermal state of the Hamiltonian (\ref{hamiltonian}), Fig.~ \ref{thermal01} depicts the negativity (solid lines), geometric discord (dashed lines) and non-locality (dotted lines) as function of $\beta$ for $m/p = 0$ (first column), $1$ (second column) and $10$ (third column), with all other parameters in correspondence with Fig.~\ref{EigenstatesTensor}.
Notice that quantum correlations are destroyed by the temperature.
In particular, the entanglement suddenly vanishes for some well-identified temperature which depends on the value of $m/p$. The non-local characteristic of quantum correlations is still more affected by the temperature effects: as the temperature increases, the state becomes local before becoming separable. In the limit of lower temperatures all correlation quantifiers tend to constant values, which depends on $m/p$, and both $\mathcal{B}$ and $\mathcal{D}$ variables match the same values. The overall dependence on $m/p$ is the same depicted in Fig.~\ref{EigenstatesTensor}, such that for $m\gg p$, quantum correlations are truly small.

\section{The role of CP symmetry transformations}

Discrete symmetries are important in the formulation of quantum field theory and in the interpretation of several phenomena in relativistic quantum mechanics. Spatial reflections, or (extrinsic) parity, P, ``geometrically mirrors'' a quantum state in a $3D$ space, and charge conjugation, C, transforms a particle state into an antiparticle state.
Both symmetries are relevant when discussing properties of fermionic Dirac-like systems. In particular, particle (intrinsic quantum number) oscillations and decay rates can be (phenomenologically) affected by the composed CP transformation, which eventually produces asymmetrical results detected by the experiments.
Under the perspective of the information theory picture discussed up to this point, the role of CP transformations on the results for {\em spin-parity} entanglement and nonlocality can be completely quantified by the CHSH inequality. CP acts as a unitary operator on both spin and intrinsic parity bi-spinor internal parameters which drive the behavior of such Dirac Hamiltonian eigenstates.
The question posed at this point is whether it may affect the bi-spinor intrinsic correlations.

To introduce the CP symmetry elements into the quantum information analysis here developed, one can conveniently introduce an auxiliary multi-space vector $\mathcal{X}$ to denote the bi-spinor parameters. For the free particle states one has $\mathcal{X} = (m, E_p)$, and for the above considered interacting states (\ref{anomalouseigenstate}) one has $\mathcal{X} = (m, \bm{p}, \bm{B}, \kappa_a, \chi_a)$. The parity transformation of a bi-spinor $\vert \psi(\mathcal{X}) \rangle$ is implemented through the unitary transformation $\hat{\mathcal{P}}$ given by
\begin{equation}
\label{Pspinor}
\hat{\mathcal{P}} \vert \psi(\mathcal{X}) \rangle = i \, \gamma_0 \vert \psi(\tilde{\mathcal{X}}) \rangle,
\end{equation}
where $\tilde{\mathcal{X}}$ is obtained by inverting the sign of all vector quantities in $\mathcal{X}$, i.e. $\tilde{\mathcal{X}} = \mathcal{X}= (m, E_p),$ for the free particle states, and $\tilde{\mathcal{X}} = (m, -\bm{p}, \bm{B}, \kappa_a, \chi_a),$ for the non-minimally coupling interacting states, where it is noticed that $\bm{B} = \nabla \times \bm{A}$ transforms as an axial-vector quantity.

Through (\ref{Pspinor}), the transformation of a density matrix $\rho$ under parity is implemented by the unitary transformation $\rho^\mathcal{P} (\mathcal{X}) = \hat{\mathcal{P}} \rho (\tilde{\mathcal{X}}) \hat{\mathcal{P}}^{-1}$, which in terms of the Fano decomposition (\ref{decomposition}) is given by
\begin{eqnarray}
\label{parity}
\rho^{\mathcal{P}}(\mathcal{X}) &=& (\hat{\sigma}_z^{(1)} \otimes \hat{I})\rho(\mathcal{X})(\hat{\sigma}_z^{(1)} \otimes \hat{I}) \\
&=&\frac{1}{4} \left[ \hat{I} + (\hat{\bm{\sigma}}^{(1)} \otimes \hat{I}_2^{(2)}) \cdot \bm{a}^{\mathcal{P}}_1(\tilde{\mathcal{X}}) + (\hat{I}_2^{(1)} \otimes \hat{\bm{\sigma}}^{(2)})\cdot \bm{a}^{\mathcal{P}}_2 (\tilde{\mathcal{X}}) + \displaystyle \sum_{i = \{x,y,z\}} (\hat{\sigma}_i \otimes \hat{\sigma}_j) \, t_{ij}^{\mathcal{P}}(\tilde{\mathcal{X}}) \right],\nonumber
\end{eqnarray}
where the parity transform of the Bloch vectors $\bm{a}^{\mathcal{P}}_{1 (2)}$ and of the correlation matrix $T^{\mathcal{P}}$ are given in terms of the original quantities as
\begin{eqnarray}
\label{ptransfproperties}
\bm{a}^{\mathcal{P}}_1(\mathcal{X}) &=& (-a_{1,x} (\tilde{\mathcal{X}}), -a_{1,y} (\tilde{\mathcal{X}}),+ a_{1,z} (\tilde{\mathcal{X}}) ), \nonumber \\
\bm{a}^{\mathcal{P}}_2(\mathcal{X}) &=& \bm{a}_2(\tilde{\mathcal{X}}), \nonumber \\
T^{\mathcal{P}} (\mathcal{X}) &=& \mbox{Diag}\{-1,-1,+1\} \,T(\tilde{\mathcal{X}}).
\end{eqnarray}

Analogously, the charge conjugation acts on a Dirac state as
\begin{equation}
\label{Cspinor}
\hat{\mathcal{C}} \vert \psi(\mathcal{X}) \rangle= i \gamma_y \vert \psi (\mathcal{X}) \rangle^*,
\end{equation}
where stars ``${*}$'' denotes the complex conjugation. In terms of density operators, the C transform of the density matrix is given by
\begin{eqnarray}
\label{chargeconjugation}
\rho^\mathcal{C}(\mathcal{X}) &=& (\hat{\sigma}_y^{(1)} \otimes\hat{\sigma}_y^{(2)} ) \rho^*(x) (\hat{\sigma}_y^{(1)} \otimes\hat{\sigma}_y^{(2)} ) = \\
&=& \frac{1}{4} \left[ \hat{I} + (\hat{\bm{\sigma}}^{(1)} \otimes \hat{I}_2^{(2)}) \cdot \bm{a}^{\mathcal{C}}_1(\mathcal{X}) + (\hat{I}_2^{(1)} \otimes \hat{\bm{\sigma}}^{(2)})\cdot \bm{a}^{\mathcal{C}}_2 (\mathcal{X}) + \displaystyle \sum_{i = \{x,y,z\}} (\hat{\sigma}_i \otimes \hat{\sigma}_j) \, t_{ij}^{\mathcal{C}}(\mathcal{X}) \right],\nonumber
\end{eqnarray}
with the Bloch vectors and the correlation matrix given by
\begin{eqnarray}
\label{ctransfproperties}
\bm{a}^{\mathcal{C}}_1(\mathcal{X}) &=& - \bm{a}_1(\mathcal{X}), \nonumber \\
\bm{a}^{\mathcal{C}}_2(\mathcal{X}) &=& - \bm{a}_2(\mathcal{X}), \nonumber \\
T^{\mathcal{C}} (\mathcal{X}) &=& T(\mathcal{X}).
\end{eqnarray}
From two-qubit operation perspective, the charge conjugation (\ref{chargeconjugation}) is a kind of spin-flip operation, which only inverts the Bloch vectors, and does not change any of the local correlation inherent properties of the state.

The combination of charge conjugation and parity, i.e. the CP transformation is thus given by
\begin{equation}
\label{CPtwoq}
\rho^{\mathcal{C} \mathcal{P}} (x) = \alpha_y \rho^*(\tilde{x}) \alpha_y = (\hat{\sigma}_x^{(1)} \otimes \hat{\sigma}_y^{(2)}) \rho^*(\tilde{x}) (\hat{\sigma}_x^{(1)} \otimes \hat{\sigma}_y^{(2)}),
\end{equation}
and the corresponding transformed Bloch vectors and correlation matrix are given by the composition of (\ref{ptransfproperties}) and (\ref{ctransfproperties}) as
\begin{eqnarray}
\label{cptransfproperties}
\bm{a}^{\mathcal{C} \mathcal{P}}_1 (\mathcal{X}) &=& (+a_{1,x} (\tilde{\mathcal{X}}),+ a_{1,y} (\tilde{\mathcal{X}}),- a_{1,z} (\tilde{\mathcal{X}}) ) \nonumber \\
\bm{a}^{\mathcal{C} \mathcal{P}}_2 (\mathcal{X}) &=& -\bm{a}_2 (\tilde{\mathcal{X}}) \nonumber \\
T^{\mathcal{C} \mathcal{P}} (\mathcal{X}) &=& \mbox{Diag}\{-1,-1,+1\} \,T(\tilde{\mathcal{X}}).
\end{eqnarray}

The C, P and CP transformations indeed act through a non-trivial sense onto the internal subsystems associated to the Dirac bi-spinors. If a Dirac bi-spinor is interpreted as a two-qubit state, C, P and CP are necessarily two-qubit operations, and as such they can modify the correlation content of a given state. When pure bi-spinors are considered, for example, in case of (\ref{anomalousstate}) with only one term in the mixture, all correlation properties are given in terms of the modulus of any of the Bloch vectors $a^2 = a_1^2 = a_2^2$. For such states one has that the parity can only change the entanglement content of the state by changing the dependence of such quantity on the parameters of the system $\mathcal{X}$. For any entanglement quantifier for pure states generically denoted by $\mathcal{E}$, one thus has $\mathcal{E}^{\mathcal{C} \mathcal{P}}(\mathcal{X}) = \mathcal{E}(\tilde{\mathcal{X}})$.

Otherwise, for mixed states, the transformations of correlations under CP are more complex, as their evaluation rely on finding specific maximum eigenvalues (c.f. Eqs.~\eqref{negativity}, \eqref{geometricdiscord} and \eqref{bellfunction}). Nevertheless, the action of CP on the nonlocality of quantum correlations can be systematically calculated. The (non)local characteristic, quantified by the Bell-CHSH function (\ref{bellfunction}), depends on the eigenvalues of $M = T^T T$.
From (\ref{cptransfproperties}), one has
\begin{equation}
M^{\mathcal{C} \mathcal{P}} (\mathcal{X})= (T^{\mathcal{C} \mathcal{P}}(\mathcal{X}))^T T^{\mathcal{C} \mathcal{P}}(\mathcal{X}) = (T(\tilde{\mathcal{X}}))^T T(\tilde{\mathcal{X}}) = M (\tilde{\mathcal{X}}).
\end{equation}
The modifications on any nonlocal character of the intrinsic correlations in Dirac bi-spinor are, therefore, uniquely driven by changes on the parameters of the state under spatial reflection.
Any changes on the internal subsystems due to CP transformations does not affect the (non)local characteristic of intrinsic quantum correlations in mixed Dirac bi-spinors.
A trivial example is identified from the free particle mixed state which depends only on two parameters: $E_p = \sqrt{p^2 + m^2}$ and $m$, where $\mathcal{X} = \tilde{\mathcal{X}}$.
Of course, the local character of such state is invariant under CP transformations. In fact, $\rho_{\mbox{\tiny free}}$ is entirely invariant under the considered CP transformation, i.e. $\rho_{\mbox{\tiny free}}^{\mathcal{C} \mathcal{P}} = \rho_{\mbox{\tiny free}}$, although the Hamiltonian itself is not (due to the mass term).

\begin{figure}[H]
\centering
\includegraphics[width= 16 cm]{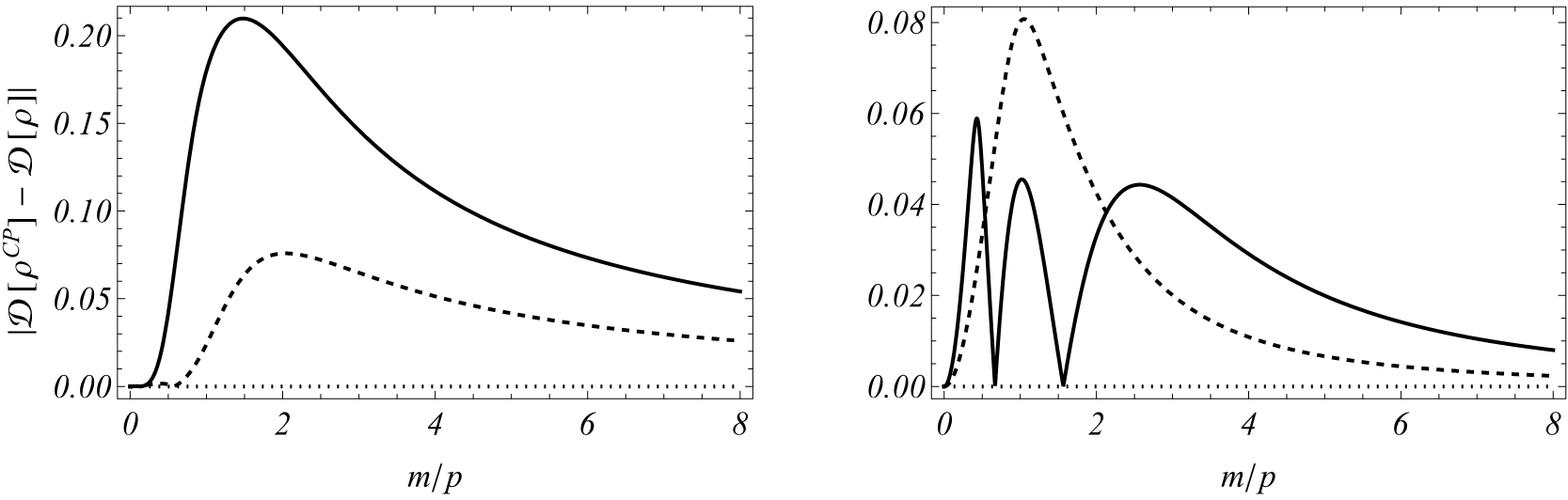}
\renewcommand{\baselinestretch}{1.0}
\caption{Absolute values of the difference between the geometric discord of the original and the CP transformed state for mixtures (c.f. Eq.~(\ref{anomalousstate})), for mixtures given by $A \rho_{00} + (1-A) \rho_{01}$ (left plot) and by $A \rho_{00} + (1-A) \rho_{11}$ (right plot). The plots are for $A = 0.1$ (solid line), $0.3$ (dashed line) and $0.5$ (dotted line). All other parameters are in correspondence with Fig.~\ref{EigenstatesTensor}. The quantum correlations in mixtures between positive and negative energy eigenstates are more robust to CP transformations. For maximal mixtures, $\rho^{\mathcal{C} \mathcal{P}}$ has the same quantum correlations of the original state.}
\label{GeometricDiscordDiff}
\end{figure}

Otherwise, for the case of an interacting dynamics, mixed states as those from (\ref{anomalousstate}) have more intricate structures. For such states, the CP transform of the Bloch vectors and of the correlation matrix (\ref{parameters}) reads
\begin{eqnarray}
\bm{a}_1^{\mathcal{C} \mathcal{P}} &=& (- a_{1x} (\mathcal{X}), a_{1y} (\mathcal{X}), a_{1z} (\mathcal{X}) ), \nonumber \\
\bm{a}_2^{\mathcal{C} \mathcal{P}} &=& (- a_{2 x}(\mathcal{X}), -a_{2 y}(\mathcal{X}), a_{2 z}(\mathcal{X})), \nonumber \\
T^{\mathcal{C} \mathcal{P}} &=& \left[ \begin{array}{ccc} t_{xx} (\mathcal{X}) & -t_{xy} (\mathcal{X}) & 0 \\ t_{yx} (\mathcal{X}) & - t_{yy} (\mathcal{X}) & -t_{yz} (\mathcal{X}) \\ t_{zx} (\mathcal{X}) & t_{zy} (\mathcal{X}) & - t_{zz} (\mathcal{X}) \end{array}\right].
\end{eqnarray}
With the above transformations one concludes that $M^{\mathcal{C} \mathcal{P}}$ has the same eigenvalues of $M$ and thus the Bell function (\ref{bellfunction}) returns the same values for both states. Therefore both states exhibit the same correlation pattern. Moreover, $\mathcal{N}[\rho^{\mathcal{C}\mathcal{P}}] = \mathcal{N}[\rho]$, i.e. the entanglement is invariant under CP for this case. The CP transformation signature is indeed manifested by the geometric discord. Fig.~ \ref{GeometricDiscordDiff} depicts the difference $\vert \, \mathcal{D} [\rho^{\mathcal{C} \mathcal{P}}] - \mathcal{D} [\rho] \, \vert$ as function of the kinematical regime $m/p$ for the mixtures $A \rho_{00} + (1-A) \rho_{01}$ (left plot) and $A \rho_{00} + (1 - A) \rho_{11}$ (right plot), for the same parameters of Fig.~\ref{EigenstatesTensor}. Mixed states composed by positive and negative energy eigenstates display quantum correlations more robust to CP transformations when compared to mixed states of positively defined energy eigenstates.
The maximal value of the difference $\vert \, \mathcal{D} [\rho^{\mathcal{C} \mathcal{P}}] - \mathcal{D} [\rho] \, \vert$ for such states is one-half of the former. For maximal mixtures, the CP transformation does not change the quantum correlations (dotted line).

Finally, for completeness, the analysis for a Hamiltonian including non-minimal coupling to an external {electric field} $\bm{E}$ could also be carry out through the substitutions
\begin{eqnarray}
\bm{B} \rightarrow \bm{E}, \hspace{0.5 cm} \chi_a \rightarrow \kappa_a, \hspace{0.5 cm} \kappa_a \rightarrow -\chi_a.
\end{eqnarray}
Because $\bm{E}$ is a vector quantity that is invertible under parity transformation, one has the parity effects parameterized by $(x, \bm{p}, \bm{E}) \rightarrow (x, -\bm{p}, -\bm{E})$. By following the same systematic approach, one proves that, for electric field interactions, not only the entanglement negativity and the Bell function locality, but also the geometric discord, are CP invariant quantities. For mixed states describing such systems, all correlations are invariant under CP, although neither the Hamiltonian nor the state are CP invariant.

\begin{figure}[H]
\centering
\includegraphics[width= 8 cm]{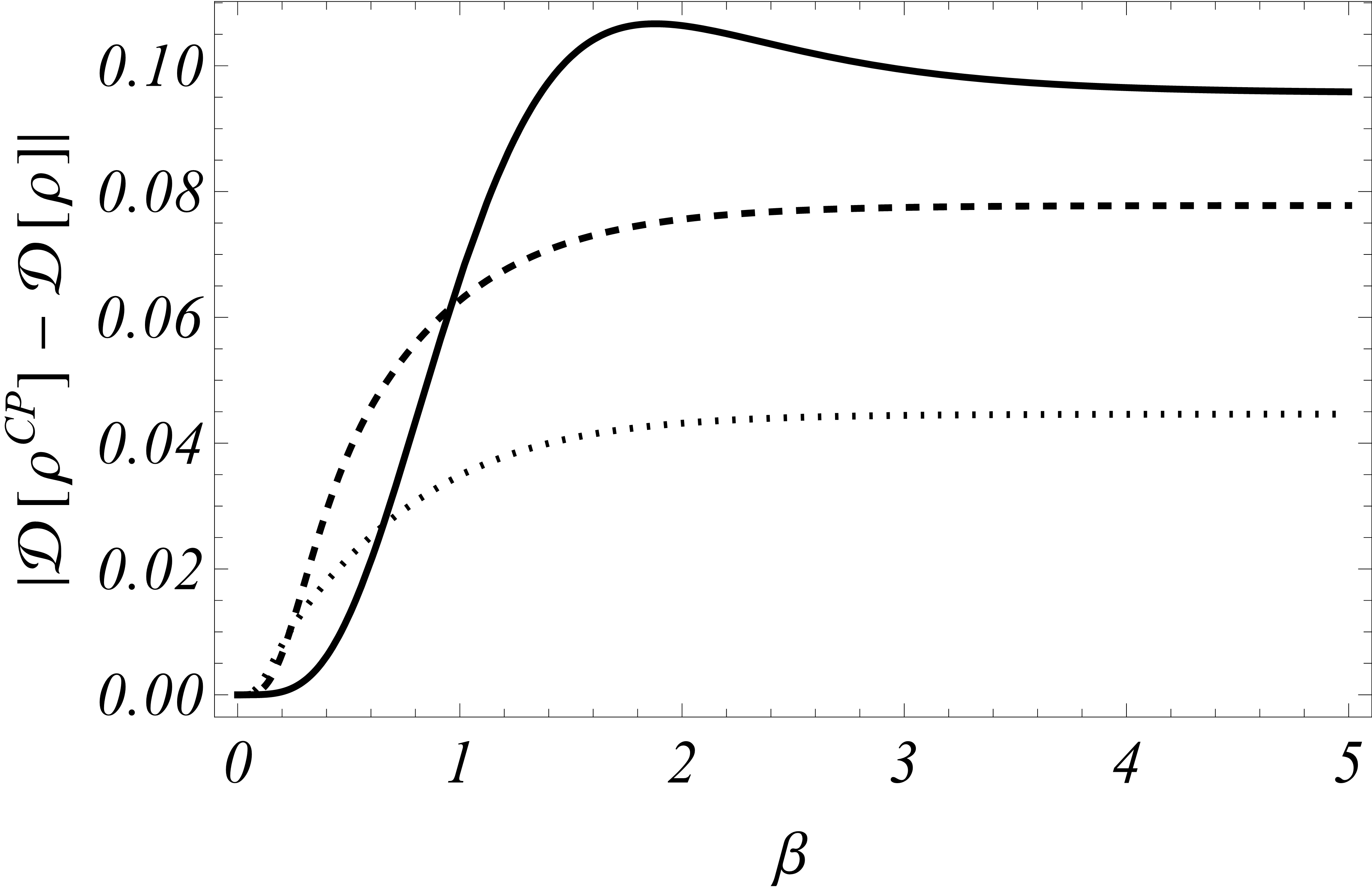}
\renewcommand{\baselinestretch}{1.0}
\caption{Absolute values of the difference between the geometric discord of the thermal state and its CP transformed state for $m/p = 1$ (solid line), $5$ (dashed line) and $10$ (dotted line) as function of $\beta = 1/T$. As generically noticed from Fig.~\ref{GeometricDiscordDiff}, the CP asymmetric effects are damped for $m \gg p$.}
\label{GeometricDiscordDiffThermal}
\end{figure}

To conclude, the effects of CP transformation on the considered thermal states from (\ref{thermal}) show some differences between quantum correlations of the CP transformed and the original states.
The effects are evinced for low temperature regimes, as depicted in Fig.~\ref{GeometricDiscordDiffThermal} for $\vert \, \mathcal{D}^{\mathcal{C} \mathcal{P}} [\rho_{\mbox{{\tiny Thermal}}}] - \mathcal{D} [\rho_{\mbox{{\tiny Thermal}}}] \, \vert$ as function of $\beta$, for $m/p = 1$ (solid line), $5$ (dashed line) and $10$ (dotted line). For lower values of $m/p$, $\vert \, \mathcal{D}^{\mathcal{C} \mathcal{P}} [\rho_{\mbox{{\tiny Thermal}}}] - \mathcal{D} [\rho_{\mbox{{\tiny Thermal}}}] \, \vert$ does not increase monotonically and exhibits a maximum value.
Therefore for such states the CP and the original states are more distinguishable (from the geometrical discord aspects) at some specific temperature value.
The damping behavior which follows $m/p$ values showed in Fig.~\ref{GeometricDiscordDiff} is also exhibited by the thermal state.

To summarize, the CP transformations modify quantum correlations in mixed Dirac states, but the nonlocal character can only be changed by a modification on the parameters of the states. In terms of two-qubit operators, the C transformation is just a spin-flip operation applied to the bi-spinor internal space and, by itself, does not change the correlations of the state. Moreover, entanglement is invariant under CP, as this operation is a globally unitary one. In general, CP asymmetries are manifested through the geometric discord. Generically, the geometric discord is the unique quantum correlation which is affected by CP.

\section{Conclusions}

The conflict between quantum mechanics and locality has suggested a profound incompatibility between quantum mechanics and relativity. Some relevant elements of such a controversial debate were discussed in the context of a typically relativistic quantum mechanical framework for Dirac(-like) systems. The relation among nonlocality, quantum correlations and quantum entanglement for mixed Dirac states were investigated under the perspective of including additional elements of charge conjugation and parity symmetries for free particle and interacting scenarios.

According to our results, free particle states prepared as a separable mixture between positive and negative helicities exhibit quantum correlations that are local, and therefore reproducible by a suitable LHV model. For non-minimal coupling scenarios, the quantum correlations in mixed states exhibit a more complex structure. For instance, the correlation between spin and intrinsic parity for mixed states composed by positive energy eigenstates tends to be more local, such that the violation of the CSHS inequality is relatively suppressed.
It is opposite to what happens for mixtures involving the superposition of positive and negative energy eigenstates for which the nonlocal correlations prevail.
The local characteristic of such mixtures are non-monotonic, different from what happens with mixtures between positively (or negatively) defined energy eigenstates, for which the locality measures always decrease for increasing values of $m/p$.
It is also worth to mention that when maximal mixtures are considered, the quantum entanglement and the additional quantum correlations are local. In this scenario, thermal effects were also quantified. In particular, one has identified that the nonlocality measure is more suppressed by temperature effects when it is compared to entanglement and quantum correlations. The thermalized mixture becomes local at temperatures lower than those required to completely destroy the entanglement.
Moreover, the geometric discord is still more resilient to temperature effects, since it vanishes only for high temperatures, when the thermal state can be approached by a maximal {\em spin-parity} mixture.

A second issue considered here is related to how CP transformations change the correlations of mixed Dirac states.
One have noticed that even states that are not invariant under CP transformations can exhibit identical profiles of quantum correlations and nonlocality.
Once specialized to the two-qubit representation of Dirac bi-spinors, the charge conjugation is just the spin flip operation implemented onto {\em spin-parity} qubits: it does not change the correlations of the bi-spinor, even if the state itself is not invariant. On the other hand, the parity symmetry acts either by operating on the combined {\em spin-parity} Hilbert space, or by changing the parameters on which the state depends. In particular, the locality content of a given mixed Dirac state can only change under CP by a changing on the parameters of the state. This fact has been exemplified by the discussion of non-minimal coupling Hamiltonians: the eigenstates are not invariant under CP, but the local characteristic of each of them, as well as its entanglement content, are invariant under CP. For such states, the effects of a CP transformation are only manifested by the results for the geometric discord. For the thermal state, the difference on quantum correlations between the CP transformed and the original state is more evinced by low temperature regimes. In addition, depending on the mass-momentum ratio, the difference does not monotonically increase with $\beta$, such that one can identify the exact temperature for which the quantum correlations are maximally distorted by CP operations.

To sum up, our results set up the framework for quantifying the local properties of Dirac bi-spinor intrinsic correlations. For some states, the quantum correlations encoded by spin and parity are typically local, and can be provided by an underlying LHV theory. More generically, the quantum discord was shown to be relevant on the characterization of how CP acts on the {\em spin-parity} correlations of such states. For example, for the case of a non-minimal coupling with an electric field, by simple substitutions one can conclude that the CP transformed states (electron and positron with opposite helicity states) shall exhibit the same local characteristic and the same degree of entanglement and quantum correlations.
As a conditional result, our preliminary analysis show that minimal electromagnetic couplings, and non- minimal coupling to electric fields does not produce quantum correlation CP violation effects.
Otherwise, non-minimal couplings with magnetic fields, as those which typically affects the content of electroweak interactions, do affect the quantum correlations driven by the geometric discord under CP transformations -- a result which claims for a deeper investigations of such correspondence between geometrical discord and CP violation.

Finally, the formalism adopted here can also be extended to the phenomenology of more feasible low energy measurable systems. For instance, the trapped ion setup can be engineered to reproduce the Dirac dynamics driven by tensor and pseudotensor potentials \cite{ MeuPRA}. The entanglement of Dirac equation solutions are then re-interpreted in terms of ionic state variables, and the intrinsic {\em spin-parity} parameters are translated into quantum numbers related the total angular momentum and to its projection onto the trapping magnetic field \cite{MeuPRA}. In this setup, the mixed state can arise as a consequence of coupling to a thermal environment, and all questions addressed here can, in principle, be measured through usual quantum optics techniques. A second physical system that can be considered is the bi-layer graphene. The Dirac equation structure of graphene is known to drive unique electronic properties, such as anomalous Hall effects, which can find applications in different areas of science and technology. More recently, it was proved that the tight-binding model for bi-layer graphene can be directly mapped into a Dirac equation which includes both tensor and pseudovector external field interactions. In this setup, some eventual transformations under CP can lead to a non-invariant local characteristic of mixed states, a challenging prospect which deserves more investigation in some future issues.

{\em Acknowledgments - The work of AEB is supported by the Brazilian Agency FAPESP (grant 17/02294-2). The work of VASVB is supported by the Brazilian Agency CAPES (grant 88881.132389/2016-1).}

\end{document}